7/25/2017

# Astronomical Observability of the Cassini Entry into Saturn

Ralph D. Lorenz

Johns Hopkins Applied Physics Laboratory

Laurel, MD 20723, USA

ralph.lorenz@jhuapl.edu

Abstract

The Cassini spacecraft will enter Saturn's atmosphere on 15th September 2017.  This event may be visible from Earth as a 'meteor' flash, and entry dynamics simulations and results from observation of spacecraft entries at Earth are summarized to develop expectations for astronomical observability.

1. Cassini End of Mission Scenario

Cassini was originally designed to perform a 4-year exploration of the Saturnian system, after arrival in 2004 and delivering the Huygens probe to Titan.  Robust design and a rich scientific return have permitted and motivated two mission extensions, first to 2010 and later to 2017. At this point, systems and instruments, although generally performing very well, are well beyond their qualification lifetimes. Electrical power has been slowly declining (restricting the number of instruments that can operate simultaneously), but most importantly the propellants for manoeuvres and attitude control have been depleted.

Planetary protection considerations require the radioisotope-powered spacecraft to be disposed of in a controlled manner that will preclude the contamination of Saturn moons that have potentially habitable environments, notably Titan and Enceladus.  As with Galileo at Jupiter in 2003, the chosen plan is to direct the spacecraft to enter the primary planet's atmosphere.

The endgame part of Cassini's orbital tour  (the "Grand Finale") was designed (e.g. Yam et al., 2009) with a series of orbits with apoapsis near Titan's orbit, permitting continued observations towards the summer solstice, and periapsis initially outside the F-ring, with a gravity assist from a  final close Titan encounter (T126, in April 2017) shifting the periapsis to within the D-ring, i.e. between the rings and the planet. These 22 final ('proximal') orbits are providing dramatic opportunities to observe the rings and the planet.  The final orbit (e.g. Bittner et al., 2016) uses a Titan encounter to adjust the final periapsis altitude to be below the Saturn cloud-tops, guaranteeing entry three days later (2017-258, 15[th] September).

Cassini will transmit data to Earth during its Saturn entry, requiring its thrusters to compensate atmospheric drag torques to maintain antenna pointing to Earth to send science data in real time at 27 kbit/s until the last possible moment.  The interval between the start of drag compensation (thrusters at 10% duty cycle) and loss of signal (thrusters at 100%, and then overwhelmed) is expected to be 1 minute. Loss of signal occurs at 2017-258T10:44:31 SCET (Spacecraft Event Time) – the one-way light time is a little over one hour, and so the event is observed at Earth at 05:08 PDT   (mission control is at the Jet Propulsion Laboratory in Pasadena, CA, USA)  or 12:08 UTC.   Unfortunately, major observatories (Chile, Hawaii) are not in view of Saturn at this time (see figure 1).

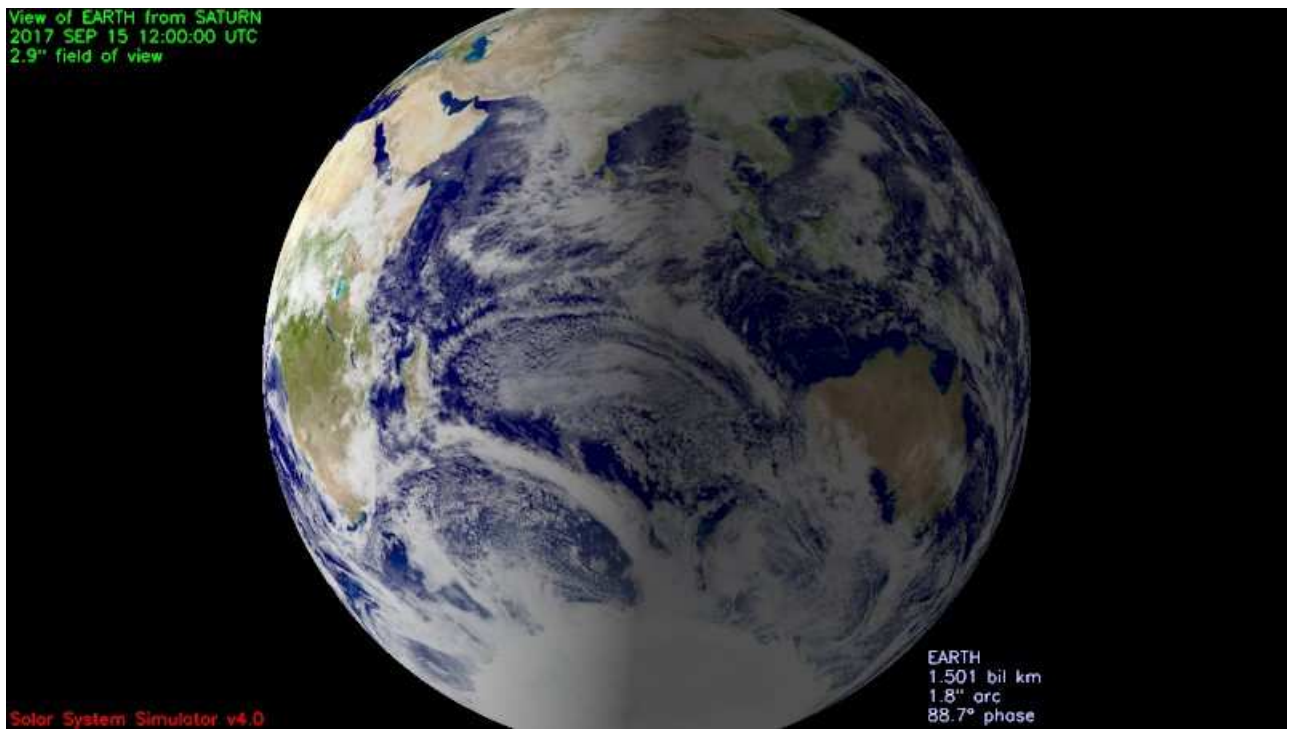

Figure 1. View of Earth from Saturn at the time of Cassini's entry.   Australia is in darkness and probably represents the best prospects for observation.  It is near sunset in India.

These times may be adjusted depending on any late updates to the spacecraft trajectory determination after the last Titan encounter, but are unlikely to change by more than a few minutes. The timetable for developing such updated estimates is not known. In any case, the trajectory and timing with respect to breakup/emission will be known post-hoc rather precisely (within a couple of seconds).

2. Entry Trajectory and Breakup

The Cassini spacecraft can be considered as a 'meteor' with a mass of 2523[1] kg and an entry speed of 34 km/s : it therefore has a kinetic energy  of some 1.4E12 J  (i.e. about one third of a kiloton 1kT=4.2E12J).

_______________________

[1] This is the dry mass : a few tens of kg of residual propellant may also be present.

The final orbit, assuming a periapsis of ~60,000km, corresponds to an entry trajectory with a flight path angle of -30 degrees defined at an atmospheric interface 1000km above the 1 bar level. This is rather steep, and will concentrate the energy dissipation into a rather short period (see table 1). The entry is computed assuming a mass of 2500kg, a fixed drag coefficient of 2.0 and a cross-sectional area of 20m2 : the atmosphere profile assumes a relative molecular mass of 2.14 and a smooth analytic temperature profile tuned to match Voyager and Cassini radio occultation data.

The spacecraft will be operating nominally until the thruster control authority is lost at an ambient atmosphere density of 9E-11 kg/m3 (a radius of 61321km). Within about 10s the aerodynamic heating will have risen to a point where gas and surfaces are heated to incandescence. The subsequent luminosity depends on the area history – large objects tend to fail mechanically and produce progressively increasing area and rapid energy deposition ('pancaking' in the bolide literature), but the dynamic pressure at which this occurs depends on the vehicle strength. Assuming 1-10 kPa (see below) this occurs between about 30 and 40s after control is lost.

Shoemaker et al., 2012 describe the observed breakup of the Hayabusa carrier spacecraft (as distinct from the entry capsule delivered by it). It was seen (their figure 9) that observable (i.e. luminous) fragments began to be shed (a suspected solar panel) at an altitude of 75km and a dynamic pressure of ~1 kPa : most fragmentation occurred at dynamic pressures of 10-30 kPa. These effective breakup strength values are appreciably lower than those typically derived for meteor breakup, 100-1000 kPa e.g. Popova et al. (2011).

The luminous efficiency u (the fraction of the instantaneous loss of energy due to drag expressed as optical emission, typically in the 300-800nm range) assumed in analysis of meteors is typically 5%. Most of this is continuum emission, although strong individual lines may be present – as discussed later. It is known that the efficiency is strongly speed-dependent, with higher speeds yielding higher efficiencies (see e.g. Revelle and Ceplecha, 2002), 34 km/s may correspond to closer to 10% efficiency. For comparison, an empirical energy-dependent efficiency is often used for terrestrial bolides, namely $u=0.12E^{0.11}$ (Brown et al., 2002) where E is energy in kilotons : for Cassini this yields u~6%.

Even without breakup (which has the effect of reducing the duration, but enhancing the intensity of energy dissipation), the peak optical emission will be of the order of 1E10 W.

Table 1 : Ballistic Entry Trajectory of Cassini assuming a nominal atmosphere profile and constant drag coefficient (Cd=2) for entry angle 30 degrees. Even without breakup, the full-width half maximum of the energy dissipation curve is only 5s.

| Time(s) | Alt(km) re 1bar | Speed (m/s) | Ambient Pressure (mb) | Density (kg/m3) | Dynamic Pressure (Pa) | Energy Dissipation (W) | |
|---|---|---|---|---|---|---|---|
| 1.00 | 983.0 | 35432.1 | 3.9E-006 | 6.11E-010 | 0.767 | 8.95E+008 | |
| 10.00 | 830.3 | 35475.6 | 5.0E-005 | 7.89E-009 | 9.93 | 9.00E+008 | |
| 21.00 | 644.3 | 35526.4 | 0.00115 | 1.78E-007 | 225. | 8.77E+008 | |
| 22.00 | 627.4 | 35530.3 | 0.00152 | 2.36E-007 | 298. | 8.68E+008 | |
| 23.00 | 610.5 | 35534.0 | 0.00202 | 3.13E-007 | 396. | 8.58E+008 | |
| 24.00 | 593.7 | 35537.3 | 0.00268 | 4.16E-007 | 525. | 8.45E+008 | |
| 25.00 | 576.8 | 35540.0 | 0.00355 | 5.51E-007 | 696. | 8.30E+008 | |
| 26.00 | 560.0 | 35542.1 | 0.00471 | 7.31E-007 | 923. | 8.15E+008 | |
| 27.00 | 543.1 | 35543.2 | 0.00624 | 9.69E-007 | 1.22E+003 | 8.04E+008 | 1kPa-breakup |
| 28.00 | 526.3 | 35543.2 | 0.00828 | 1.29E-006 | 1.62E+003 | 8.06E+008 | starts ? |
| 29.00 | 509.5 | 35541.5 | 0.0110 | 1.70E-006 | 2.15E+003 | 8.36E+008 | |
| 30.00 | 492.6 | 35537.8 | 0.0145 | 2.26E-006 | 2.85E+003 | 9.21E+008 | |
| 31.00 | 475.9 | 35531.2 | 0.0193 | 2.99E-006 | 3.78E+003 | 1.09E+009 | |
| 32.00 | 459.1 | 35520.9 | 0.0255 | 3.96E-006 | 5.00E+003 | 1.37E+009 | |
| 33.00 | 442.3 | 35505.7 | 0.0338 | 5.25E-006 | 6.62E+003 | 1.80E+009 | |
| 34.00 | 425.5 | 35484.0 | 0.0447 | 6.95E-006 | 8.75E+003 | 2.41E+009 | |
| 35.00 | 408.8 | 35453.8 | 0.0592 | 9.20E-006 | 1.16E+004 | 3.23E+009 | 10kPa |
| 36.00 | 392.1 | 35412.3 | 0.0784 | 1.22E-005 | 1.53E+004 | 4.34E+009 | |
| 37.00 | 375.4 | 35356.0 | 0.104 | 1.61E-005 | 2.01E+004 | 5.81E+009 | |
| 38.00 | 358.8 | 35280.4 | 0.137 | 2.13E-005 | 2.65E+004 | 7.74E+009 | |
| 39.00 | 342.2 | 35179.4 | 0.181 | 2.81E-005 | 3.47E+004 | 1.02E+010 | |
| 40.00 | 325.7 | 35045.4 | 0.238 | 3.70E-005 | 4.54E+004 | 1.35E+010 | |
| 41.00 | 309.3 | 34868.9 | 0.314 | 4.87E-005 | 5.92E+004 | 1.76E+010 | |
| 42.00 | 293.0 | 34637.8 | 0.412 | 6.40E-005 | 7.68E+004 | 2.28E+010 | |
| 43.00 | 276.9 | 34337.4 | 0.541 | 8.63E-005 | 1.02E+005 | 2.93E+010 | |
| 44.00 | 260.9 | 33937.2 | 0.721 | 0.000125 | 1.44E+005 | 4.01E+010 | |
| 45.00 | 245.3 | 33370.5 | 0.981 | 0.000185 | 2.06E+005 | 5.61E+010 | |
| 46.00 | 230.0 | 32564.2 | 1.36 | 0.000278 | 2.94E+005 | 7.78E+010 | |
| 47.00 | 215.2 | 31423.9 | 1.91 | 0.000419 | 4.14E+005 | 1.05E+011 | |
| 48.00 | 201.2 | 29844.5 | 2.69 | 0.000632 | 5.63E+005 | 1.37E+011 | |
| 49.00 | 188.2 | 27742.0 | 3.79 | 0.000941 | 7.24E+005 | 1.65E+011 | |
| 50.00 | 176.5 | 25112.6 | 5.25 | 0.00136 | 8.60E+005 | 1.80E+011 | Peak Heating |
| 51.00 | 166.1 | 22082.3 | 7.07 | 0.00190 | 9.29E+005 | 1.74E+011 | |
| 52.00 | 157.2 | 18900.3 | 9.20 | 0.00255 | 9.10E+005 | 1.49E+011 | |
| 53.00 | 149.8 | 15848.8 | 11.5 | 0.00326 | 8.19E+005 | 1.15E+011 | |

3. Spectroscopic Observation of Previous Spacecraft Entries

The character of the emission is not well-estimated for Cassini at this time, although sophisticated coupled flow-chemistry-radiation models have been used in the past, and have given some success for predicting emission spectra and intensity histories (e.g. Boyd et al., 2013). Typically, line emissions are superposed on a black body curve of 3500-10000K.

A number of controlled spacecraft entries have been observed in the last decade or so (e.g. Winter, 2014). These include Stardust, Hayabusa, and the Ariane Transfer Vehicle (ATV) Jules Verne, returning from a comet, an asteroid, and the International Space Station, respectively. In addition to having entry at a known place and time, these missions have been supported by mobile observatories set up near the landing sites (Stardust – Utah, USA; Hayabusa – Woomera, Australia) or by an airborne observatory over the ocean (ATV, and the space debris object WT1190F, Jenniskens 2016).

The Stardust entry in some respects is not unlike a conventional meteor entry, in that the vehicle was a compact shape with appreciable strength (the sample return capsule was designed to survive entry, protected by a carbon heat shield). An extensive literature exists on observing and interpreting meteor entries. The Hayabusa and ATV entries are perhaps more representative of Cassini in terms of structural strength and material composition.

From an entry dynamics standpoint, Cassini is rather more energetic than all these cases – its entry velocity of 34 km/s being much larger. Cassini is not quite as massive as the ATV, but is rather larger than Hayabusa or Stardust. The entry angle is also rather steep, although offsetting this is the fact that the scale height of the Saturn atmosphere (with a low molecular weight) is larger, spreading the energy dissipation over a longer scale.

Spectroscopy of the Stardust entry (Steinback et al., 2010) indicated emission lines of sodium and potassium (presumably traces evolved from thermal paint on the heat shield; sodium is also a component of the glass used in printed circuit boards) as well as oxygen.

Near-IR nitrogen emission lines (1010, 1052m 1072nm) were seen in the Hayabusa capsule entry (Snively et al., 2014), as well as a 1069nm emission from carbon from the heatshield. Other observations in the visible (Jenniskens et al., 2012) showed lines of Hydrogen, Oxygen and Nitrogen, as well as Sodium, Potassium and Calcium.

The more massive ATV Jules Verne entry saw (Snively et al., 2011; Winter 2014) a range of emissions that included Magnesium, Barium, Zinc, Aluminium Oxide, as well as Nickel, Copper, Chromium and Titanium : these elements are typical components of spacecraft paints and alloys. Emission tentatively identified as lithium from a battery was noted at the end of entry. As with Hayabusa, sodium and potassium were seen (although only trace spacecraft constituents, their high ionization efficiency causes prominent emission) as well as C-N emission from the atmosphere.   Model spectra can be seen at http://atv.seti.org/preflightpredictions.jpg  and are discussed in detail in Boyd et al. (2013).

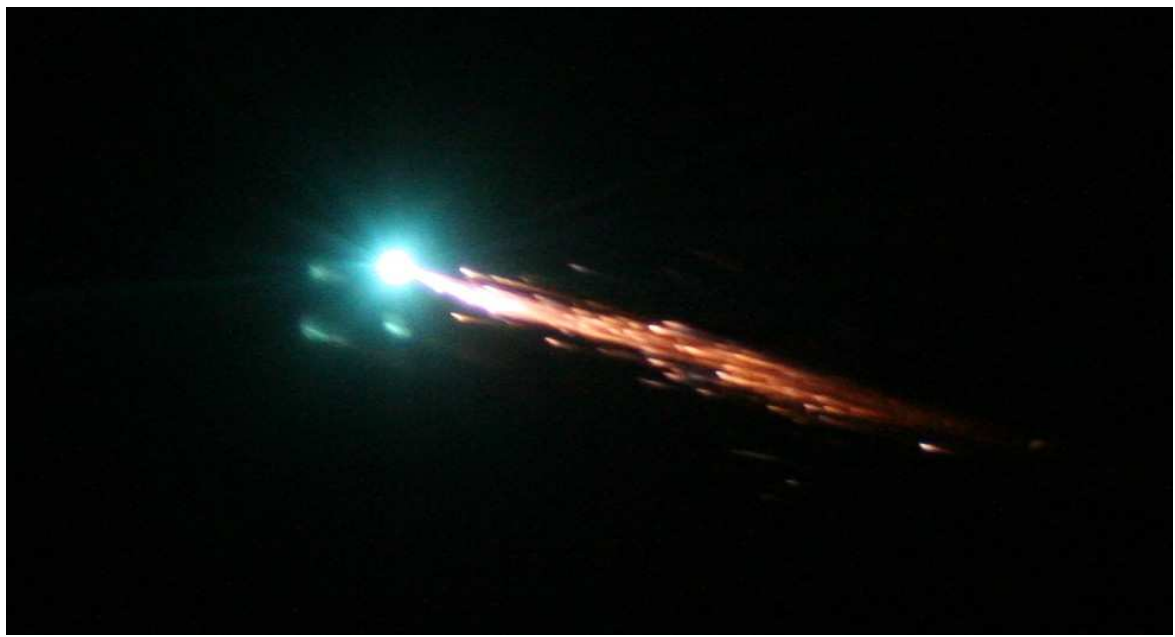

Figure 2.  Image from NASA DC-8 observatory aircraft showing the emission from the Ariane Transfer Vehicle Jules Verne entry. Note the different color of the glowing fragments from the main vehicle (ESA photo).

For Cassini's entry into Saturn, the spacecraft components should be largely similar, although Cassini does not have a battery so no prominent lithium emission should be expected.  A major difference, of course, is in the atmospheric composition, with no oxygen. On the other hand, hydrogen and perhaps helium emission lines should be much more prominent ; Saturn's atmosphere has traces of ammonia and methane, so C and N emissions are possible, but probably less prominent than at Earth.

It may be noted that an effort was made to observe the Huygens entry at Titan in 2005 (Lorenz et al., 2006) but was unsuccessful  (the most promising instrument applied, the Space Telescope Imaging Spectrometer STIS unfortunately failed before the entry).  Huygens was a much less energetic event, however  (300kg, 6 km/s -  so ~250 times lower energy than Cassini).

## 4. Giant Planet Transient Observations

The Cassini entry, as a brief optical emission on the dayside, has some similarities with attempts to detect other optical transients such as lightning or bolide entries.  In addition to the famous series of large comet fragment entries at Jupiter seen in 1994 after the breakup of comet Shoemaker-Levy/9 (where only the plumes after impact were seen from Earth), several small bolide entries have been observed at Jupiter (e.g. Hueso et al., 2013). Remarkably, these three bolides have been observed by amateur astronomers using telescopes of aperture 12.5 to 37cm with video cameras or similar CCDs operating at frame rates of 15-60 Hz. These observations have generated lightcurves of the fireballs, lasting 1.4-1.9 seconds. The flashes have optical energies of 3-32 E13 J, and correspond to 100-1000 ton bolides (see table 2).

A 2000kg spacecraft, with an entry speed of 34 km/s, is a considerably weaker event, and of course Saturn is twice as far away as Jupiter. Nonetheless, adequate light to make a detection should be possible with a reasonable aperture and perhaps a longer exposure time (a ten-point lightcurve would give useful definition of the energy deposition). The background surface brightness of Saturn is a little lower, and contrast could be improved further by using a filter to isolate wavelengths where Saturn's albedo is low – notably in methane absorption bands such as that at 889nm.

The particular challenge of detecting transients on the bright dayside of a planet has been discussed by Luque et al. (2015) who describe useful statistical techniques to discriminate sought events (in this case lightning) from cosmic ray hits on the detector. Here, the event appears only in a single pixel in a single frame in a sequence : much more robust and sensitive statistical detection schemes exist for events which may last several frames.

5. Conclusions

The Cassini entry will cause a fireball with an estimated optical energy of the order of 5-10E10 J : this is two orders of magnitude fainter than unanticipated fireballs detected at Jupiter with amateur telescopes and imaging with conventional color filters. Imaging at wavelengths where Saturn is faint (UV, and methane bands) with 1m-class telescopes appears to offer prospects that the Cassini entry might be detected : the known location and time of entry will make a search easier. Spectroscopy also holds significant promise, and previous spacecraft entries give indications of possible emission lines. Observation of the entry will confirm the disposal of the spacecraft : if time-resolved spectroscopic lightcurves can be obtained, these would be interesting for comparison with aerothermochemical models of giant planet entries by bolides and atmospheric probes.

Table 2 :  Spacecraft Entries and Comparable Events

| Entry | Diameter (m) | Mass (kg) | Speed (km/s) | Angle (deg) | Kinetic Energy (J) |
|---|---|---|---|---|---|
| Stardust | | 46 | 12.9 | -8.2 | 3.8E9 |
| Hayabusa (Carrier) | ~2 | 380 | 12.2 | -12.3 | 2.8E10 |
| Hayabusa (Capsule) | 0.4 | 16 | 12.2 | -12.3 | 1.3E9 |
| ATV-1 Jules Verne | 4.5 | 1.3E4 | 7.6 | -1.4 | 3.7E11 |
| WT1190F | ~1* | 700* | 11 | | 4.2E10 |
| Cassini | 5-10 | 2500 | 34 | -30 | 1.4E12 |
| Jupiter Bolide | 5-18 | 1E5* | 60** | 45** | 2-17E14 |

* Estimated  ** Assumed